\documentclass[12pt]{article}
\usepackage{float}
\pdfoutput = 1
\usepackage[utf8]{inputenc}
\usepackage{epsfig,color}
\usepackage{graphics}
\usepackage{graphicx} 
\begin{document}
\date{Today}
\title{{\bf{\Large   The Effect of  Modified Dispersion Relation on Dumb Holes }}}

\author{{\bf {\normalsize Abhijit Dutta}$^{a,b}$
\thanks{dutta.abhijit87@gmail.com,}},\,
{\bf {\normalsize Sunandan Gangopadhyay}$^{c,d}
	$\thanks{sunandan.gangopadhyay@bose.res.in,~sunandan.gangopadhyay@gmail.com,sunandan@associates.iucaa.in}},\,
{\bf {\normalsize Sebastian Bahamonde}$^{e}
		$\thanks{sebastian.beltran.14@ucl.ac.uk }},\,
{\bf {\normalsize Mir Faizal}$^{f,g}
	$\thanks{mirfaizalmir@googlemail.com  }}\\
$^{a}${\normalsize Department of Physics, Kandi Raj College, Kandi, Murshidabad 742137, India }\\[0.2cm]
$^{b}${\normalsize Department of Physics, West Bengal State University, Barasat, Kolkata 700126, India}\\[0.2cm]
$^{c}${\normalsize Department of Theoretical Sciences, S. N. Bose National Centre for Basic Sciences,}\\[0.2cm]
{\normalsize JD Block, Sector III, Saltlake, Kolkata 700106, West Bengal, India }\\ [0.2cm]
$^{d}${\normalsize Visiting Associate in Inter University Centre for Astronomy \& Astrophysics, Pune, India}\\[0.1cm]
$^{e}${\normalsize Department of Mathematics, University College London,
}\\
{\normalsize Gower Street, London, WC1E 6BT, UK}\\ [0.2cm]
$^{f}${\normalsize Department of Physics and Astronomy,
}\\
{\normalsize University of Lethbridge,
	Lethbridge, Alberta, T1K 3M4, Canada}\\[0.2cm]
$^{g}${\normalsize Irving K. Barber School of Arts and Sciences,}\\
{\normalsize University of British Columbia - Okanagan, 3333 University Way}\\
{\normalsize Kelowna, British Columbia V1V 1V7, Canada}
}
\date{}

\maketitle

\begin{abstract}
In this paper, we will deform the usual energy-momentum dispersion relation of a photon gas at an intermediate scale 
between the Planck and
electroweak 
scales. We will demonstrate that such a deformation can have non-trivial  effects on  the physics of dumb holes. 
 So, motivated by the physics of dumb holes, we will first analyse the effect of such a deformation on
 thermodynamics. 
 Then we observe that the velocity of sound also gets modified due to such a modification 
 of the thermodynamics. This   changes the position of the horizon of a dumb holes, and  the analogous Hawking radiation 
 from a dumb hole. Therefore, dumb holes can be used to measure the deformation of the usual energy-momentum 
 dispersion relation. 
\end{abstract}


 \section{Introduction}

In this paper, we will analyze the deformation of the usual energy-momentum dispersion relation. 
However, instead of deforming the usual energy-momentum dispersion relation at Planck scale, 
we will deform it at  an intermediate scale between the Planck and 
electroweak scales. Such a deformation will modify the physics at energy scales much lower than the 
Planck scale. We will demonstrate that it will modify the physics of a dumb hole, as it will 
modify the velocity of sound. So, we will use this modified physics of dumb hole to test 
the deformation  of the usual energy-momentum dispersion relation. 
We will study and propose a test for the deformation of the usual energy-momentum dispersion relation, 
as such a deformation is well motivated from various different theoretical considerations 
\cite{Iengo,Arkani,Gripaios,Alfaro,Belich}.
In fact, as the Lorentz symmetry fixes the form of the usual energy-momentum dispersion 
relation, so the usual energy-momentum dispersion relation will get deformed by 
the breaking of the Lorentz symmetry. Even though the Lorentz symmetry is one of the 
most important symmetries in nature, there are strong indication from various approaches to quantum gravity 
that it might only  be broken in the UV limit.   It has 
been observed that such 
deformation of the usual energy momentum dispersion relation in the UV limit (due to the breaking of Lorentz symmetry) 
occurs in the discrete spacetime \cite{Hooft}, spacetime foam \cite{Ellis},
spin-network in loop quantum gravity (LQG)
\cite{Gambini}, non-commutative geometry
\cite{FaizalMPLA,Carroll},   ghost condensation
\cite{FaizalJPA}, Horava-Lifshitz gravity \cite{h1, h2} and in some modified teleparallel theories of gravity \cite{Bahamonde:2015zma,Bahamonde:2017bps,Bahamonde:2016kba,Bahamonde:2015hza}. The existence of a modified energy-momentum dispersion relation has lead to the development of doubly special relativity (DSR)
\cite{13}, where non-linear Lorentz transformation are used to incorporate  a maximum energy scale in the theory.
The  generalization of such a theory to curved spacetime has led to the development of  gravity's rainbow \cite{gr}. 

It is possible for the usual energy-momentum dispersion to get deformed in string theory due the  breaking of Lorentz 
symmetry, which can occur because of  an
unstable perturbative string vacuum \cite{58}.  In string field theory, the tachyon field can have the wrong sign for 
its mass squared. This makes the   perturbative
string vacuum  unstable. This theory is ill-defined when the vacuum expectation value of the tachyon field is   infinite. 
However, it is possible to obtain a finite negative 
vacuum expectation value for  the
tachyon field. This will break the Lorentz symmetry as it will make the 
coefficient of the quadratic term for the massless vector field to become 
nonzero and negative. It is known that supergravity theories can be obtained as 
the low energy effective action for string theory. However, it 
is possible to construct a gravitational version of the
Higgs mechanism in supergravity theories, and this can also break the Lorentz symmetry in string theory 
\cite{59}. It may be noted that black branes in type IIB string theory have also been used to 
break the Lorentz symmetry in string theory  \cite{60}. This is done by analysing the  moduli stabilization
using a KKLT-type moduli potential in  
type IIB warped flux compactification. It has been observed that a   
Higgs phase for gravity   exists when   all moduli are stabilized.
This breaks the Lorentz symmetry in type IIB string theory. 
Compactification in string field theory has also been used for 
breaking the Lorentz symmetry in string theory \cite{61}. Thus, there is a  strong motivation to study 
the modified energy-momentum dispersion even in  string theory. 

Even though there is a strong motivation to study the deformation of the usual energy-momentum dispersion 
relation, such a deformation is usually studied at Planck scale, and so it cannot be tested using low energy 
experiments. However, it has been argued that quantum gravitational effects can occur at an 
intermediate scale between the Planck and electroweak scales, and such quantum gravitational effects can have 
low energy consiquences \cite{sa, sa12}.  
The  quantum gravitational effects  at such an intermediate scale occur due to the generalized 
uncertainty principle \cite{sa, sa12}, 
and generalized uncertainty principle is related to deformed energy-momentum  dispersion relation \cite{md01, md02},
so the deformation of the usual usual energy-momentum dispersion relation can also occur at such an 
intermediate scale. 
In fact, as  quantum gravitational effects are one of the main motivation to study Lorentz symmetry breaking, 
it is possible that the Lorentz symmetry can break at an intermediate scale between the Planck and 
electroweak scales. In this case,   the scale at which the Lorentz symmetry breaks will be bounded by the 
present day experimental results.  In fact, 
cosmic rays can be used for analyzing the breaking of Lorentz symmetry,  as they can be 
used to set upper limits on the energy at which quantum gravitational effects can occur. It is possible to use the 
Greisen-Zatsepin-Kuzmin limit (GZK limit)  to argue for the existence of a bound on the deformation of the 
usual energy-momentum dispersion relation    
\cite{Greisen,Zatsepin}.  It may be noted that even the  Pierre Auger Collaboration and the High Resolution
Fly's Eye (HiRes) experiment have reconfirmed earlier results of
the GZK cutoff, thus suggesting the deformation of the usual energy-momentum
dispersion relation in the UV limit \cite{Abraham}. 
The   hard spectra from gamma-ray 
burster's also suggests that the usual energy-momentum dispersion relation will get deformed in 
the UV limit~\cite{AmelinoCamelia:1997gz}. 
So, these experimental results can be used to obtain a bound on  the scale at which such a 
UV modification of the  usual energy-momentum dispersion relation can occur.

Now as the usual energy-momentum dispersion relation can be deformed at an intermediate scale between 
the Planck and electroweak scales, it is possible to  test  such effects. 
Therefore, in this paper, we propose to test such a deformation using the 
modification to  velocity of sound in a photon gas produced from such a deformation. 
The velocity of sound in the  gas of photons depends on the 
thermodynamic theory of the photon gas \cite{sound}, and the deformation of the usual energy-momentum 
dispersion relation will also deform the thermodynamics of this system \cite{hight, ther2,ther,ther1,ther0}. 
We will explicitly calculate this change in the velocity of sound because of a deformation 
of the energy-momentum dispersion relation. 

Such a correction to the velocity of sound can be measured using a dumb hole, as the 
horizon of  phonons in a dumb hole depends on the local  velocity of sound \cite{d1, d2}, 
and so a change in the velocity 
of sound, which change the geometry of a dumb hole.  It may be noted that such  sonic black holes have   
been studied in  Bose-Einstein condensates 
\cite{cobe, cobe1, conden}.
The Hawking radiation from such sonic black holes has   been studied \cite{hawk}. It was demonstrated that 
for such sonic black holes, quasi-particle are   radiated with a thermal distribution, and this distribution is in 
exact agreement with the distribution produced by the Hawking radiation.
The gravity dual of dumb holes has also been analysed 
using the fluid-gravity correspondence \cite{Das:2010mk}.
As the physics of a dumb holes  depends on the geometry of a dumb hole, which in turn depends on 
the local velocity of sound, it is possible to study the corrections to the physics of a dumb hole 
from the corrected velocity of sound. 
This can be measured experimentally, and so it can be used to test the deformation of 
the energy-momentum dispersion relation.

\section{Modified Dispersion Relation}
As the velocity of sound depends on the thermodynamics of the system, we will first have to analyze the deformation 
of the thermodynamics produced by the deformation of the usual energy-momentum dispersion relation. In this section, we will analyze the effect of this deformation on the partition function.
It may be noted that the  deformation of the thermodynamics of a 
 photon gas from the deformed energy-momentum 
 dispersion relation has already been studied 
 (see~\cite{hight, ther2,ther,ther1,ther0}). However, this has been studied using a modified dispersion relation proposed 
  	by Maguejo and Smolin  (MS) \cite{Magueijo-smolin, Magueijo-smolin1}. The MS deformation of the usual 
  	energy-momentum dispersion relation is given by $E^2 -p^2 = m^2 (1 - \kappa^{-1} E)^2 $, where $\kappa$ 
  	is the deformation parameter. 
 In this paper, we will use a  different   	form of the modified dispersion relation (MDR).  This dispersion relation is given by  \cite{Alae_Brand_Mag,  amelino-camelia-Mazid} 
  	\begin{eqnarray}
E^2 = p^2 (1+\lambda E)^2+m^2\,,
\label{dispersion relation}
\end{eqnarray} 
 where $\lambda$ is the deformation parameter. Note that we used natural units $c=h=1$.  It may be noted that  the thermodynamics in a 
  	high temperature limit of this modified dispersion relation has been studied in \cite{hight}. In the following, we will analyse the effect of 
  this  deformation on the thermodynamics, without taking any such high temperature limit.
Furthermore,  we will assume that 
this modification 
occurs at an intermediate scale between the Planck and electroweak scales, 
 so  this can have interesting consequences for dumb holes. 
Now for photon gas $(m=0)$, Eq.~(\ref{dispersion relation}) becomes
\begin{eqnarray}
E = p(1+\lambda E)\,.
\label{deformed dispersion for photon}
\end{eqnarray} 
Parameter $\lambda$ is related with maximum momentum $p_{\rm max}= {1}/{\lambda }$. 
It may be noted that this maximum momentum is related to a Planckian  momentum 
in the usual approach, and in the limit
$\lambda \rightarrow 0$, we obtain the usual energy-momentum dispersion relation 
$E = p $. 
 
\noindent The usual single particle  partition function  $Z_{1}(T,V)$  can be written as 
\begin{eqnarray}
Z_{1}(T,V) = 4\pi V \int_{0}^{\frac{1}{\lambda}} p^2 e^{-\beta E} dp\,.
\label{SR partition fn.}
\end{eqnarray}
where $\beta = 1/(k_{B}T)$, $k_{B}$ is the Boltzmann constant and $T$
is the temperature of the particle. By solving Eq. (\ref{deformed dispersion for photon}) for the momentum $p$, 
one gets $p=E/(1+\lambda E)$. Now by replacing that expression in the above integral,
 the partition function for one single particle becomes
\begin{eqnarray}
Z_{1}(T,V) = 4\pi V \int_{0}^{\infty} \frac{E^2 e^{-\beta E}}{(1+\lambda E)^4}dE\,.
\label{SR partition fn2.}
\end{eqnarray}
The above integral cannot be computed analytically.
However, it may be noted that this partition function has a intrinsic cut-off at the order $\lambda$.
Thus one can rewrite the above integral as follows
\begin{eqnarray}
Z_{1}(T,V)&=& 4\pi V\left(\int_{0}^{\frac{1}{\lambda}}\frac{E^2 e^{-\beta E}}{(1+\lambda E)^4}dE+ \int_{\frac{1}{\lambda}}^{\infty}\frac{E^2 e^{-\beta E}}{(1+\lambda E)^4}dE\right)~.
\label{Z}
\end{eqnarray}

\noindent  We proceed to compute the integrals in the above expression. Since $\lambda E \leq 1$ in 
the first integral, by a power series expansion
in $\lambda E$, the first integral in Eq.~(\ref{Z}) can be written as
\begin{eqnarray}
\int_{0}^{\frac{1}{\lambda}}\frac{E^2 e^{-\beta E}}{(1+\lambda E)^4}dE = \int_{0}^{\frac{1}{\lambda}}E^2 e^{-\beta E}dE - 4\lambda \int_{0}^{\frac{1}{\lambda}}E^3 e^{-\beta E}dE + 10 {\lambda}^2 \int_{0}^{\frac{1}{\lambda}}E^4 e^{-\beta E}dE +\mathcal{O}(\lambda^3).
\label{900}
\end{eqnarray}
The integrals appearing in the above expression have the general form
\begin{eqnarray}
I_n = \int_{0}^{\frac{1}{\lambda}}E^n e^{-\beta E}dE\,,\quad n=2,3,4.
\end{eqnarray}
 The integrals can be computed, and up to second order in $\lambda$, we obtain  
\begin{eqnarray}
\int_{0}^{\frac{1}{\lambda}}{E^2 {(1-4\lambda E + 10 {\lambda}^2 E^2)} e^{-\beta E}}dE &=&-\frac{e^{-\frac{\beta}{\lambda}}}{\beta^3}\left(7\frac{\beta^2}{\lambda^2}+30\frac{\beta}{\lambda}+98+216\frac{\lambda}{\beta}+240\frac{\lambda^2}{\beta^2}
\right)\nonumber\\
&&+\frac{2}{\beta^3}\left(1-\frac{12\lambda}{\beta}+\frac{120\lambda^2}{\beta^2}\right)+\mathcal{O}(\lambda^3).\nonumber\\
\label{Z_1st_int}
\end{eqnarray}
For the second integral in Eq.~(\ref{Z}), $\lambda E \geq 1$. Hence, we carry out 
a power series expansion in $1/(\lambda E)$, which upto second order in $\lambda$ yields
\begin{eqnarray}
\frac{1}{\lambda^4}\int_{\frac{1}{\lambda}}^{\infty}\frac{e^{-\beta E}}{E^2}\left(1-\frac{4}{\lambda E}+\frac{10}{\lambda^2 E^2}\right) dE &=&\frac{e^{-\frac{\beta}{\lambda}}}{\beta^3}\left(7\frac{\beta^2}{\lambda^2}-30\frac{\beta}{\lambda}+158-984\frac{\lambda}{\beta}+7040\frac{\lambda^2}{\beta^2}
\right)\nonumber\\
&&+\mathcal{O}(\lambda^3)\,.\label{Z_2nd_int}
\end{eqnarray} 
By substituting the expansions of the integrals given in Eq.~(\ref{Z_1st_int}) and Eq.~(\ref{Z_2nd_int}) into Eq.~(\ref{Z}), 
we obtain  the   partition function (up to second order in $\lambda$), 
\begin{eqnarray}
Z_1(T, V)&=& 4\pi V\left[\frac{2}{\beta^3}\left(1-\frac{12\lambda}{\beta}+\frac{120\lambda^2}{\beta^2}\right)-60\frac{e^{-\frac{\beta}{\lambda}}}{\beta^3}\left(\frac{\beta}{\lambda}-1+20\frac{\lambda}{\beta}-114\frac{\lambda^2}{\beta^2}
\right)\right]+\mathcal{O}(\lambda^3).\nonumber\\
\label{single paricle partition func}
\end{eqnarray}

\noindent Finally, for a $N$ classical particle system, the partition function $Z_{N}(V,T)$,  can be written as 
\begin{eqnarray}
Z_{N}(T, V) = \frac{1}{N!}[Z_{1}(T, V)]^N\,.
\label{N particle function}
\end{eqnarray} 
Thus, we have obtained the expression for the partition function for the deformed photon gas.
Now we will be able to analyse the effect of such a 
deformation on the thermodynamics of a photon gas. 
\section{Thermodynamics  of  Photon Gas}
In this section, we will analyse certain aspects of thermodynamics of photon gas by asuming a modified 
energy-momentum dispersion relation \cite{ther}. 
The thermodynamics of a photon gas depend on the usual energy-momentum dispersion relation, so a deformation of
the usual energy-momentum dispersion relation 
will deform the thermodynamics of the photon gas. It may be noted that such a modified   thermodynamics
has been studied in the 
  	high temperature limit \cite{hight}. However, in this section, we will analyse the
  	deformed thermodynamics without using any 
  	such limit. 
  	This will be important, as we will use this deformed thermodynamics to dumb holes.
  	However, dumb holes are usually obtained in
  	Bose-Einstein condensates 
  \cite{cobe, cobe1}, and these condensates become unstable at high temperatures. Thus, 
  we will first have to analyse the effect of such 
  a deformation without taking any high temperature limit.
The free energy of the photon gas is given by 
\begin{eqnarray}
F &=& - k_BT \ln (Z_{N}(T, V))\,. 
\end{eqnarray}
Now, by using Eqs.~(\ref{single paricle partition func}) and (\ref{N particle function}),  we obtain
\begin{eqnarray}
F &=&-Nk_BT\Big[1+\ln\Big\{\Big(\frac{4 \pi Vk_B^3T^3}{N}\Big)Q(\lambda,T)\Big\}\Big]+\mathcal{O}(\lambda^3)\,,
\label{free energy}
\end{eqnarray}  
where for simplicity we have defined the function
\begin{eqnarray}
Q(\lambda,T)&=&2-24\lambda  k_BT+240\lambda^2  {k_B}^2 T^2\nonumber\\&&
-60 e^{-\frac{1}{\lambda k_BT}}\Big(\frac{1}{\lambda k_BT}-1+20\lambda k_BT-114\lambda^2  {k_B}^2 T^2
\Big)\,.
\end{eqnarray}
The above quantity was defined to avoid long equations since it will also appear in the next thermodynamics quantities. \\
The expression of the free energy can be used to calculate the pressure,   
\begin{eqnarray}
P= -\left(\frac{\partial F}{\partial V}\right)= \frac{Nk_BT}{V}\,,
\label{pressure}
\end{eqnarray}
so that the equation of state $ PV = Nk_BT$ is not deformed by the deformation of the energy-momentum dispersion relation.

\begin{figure}[H]
	\centering
	\includegraphics[scale=0.7]{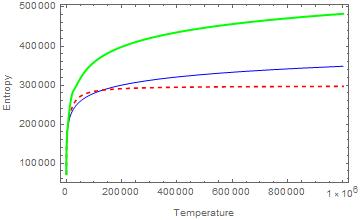}
	\caption{\textbf{Entropy $S$ as a function of the Temperature $T$ for three cases}: Special Relativity (SR), Magueijo-Smolin (MS) model 
	and Modified Dispersion Relation (MDR). Here upper curve (green line) represents MDR model, middle curve (blue line) 
	represents SR and lower curve (red line) represents MS model. }
	\label{Fig1}
\end{figure}

\begin{figure}[H]
	\centering
	\includegraphics[scale=0.7]{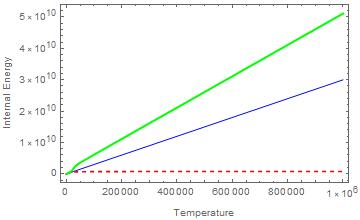}
	\caption{\textbf{Internal energy as a function of the Temperature $T$ for three cases}: Special Relativity (SR), Magueijo-Smolin (MS) model 
	and Modified Dispersion Relation (MDR).   }
	\label{Fig2}
\end{figure}

\begin{figure}[H]
	\centering
	\includegraphics[scale=0.7]{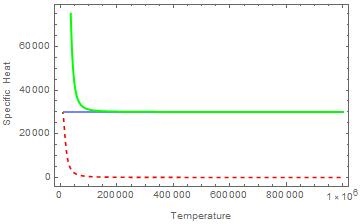}
	\caption{\textbf{Specific heat as a function of the Temperature $T$ for three cases}: Special Relativity (SR), Magueijo-Smolin (MS) model 
	and Modified Dispersion Relation (MDR).   }
	\label{Fig3}
\end{figure}

\noindent The entropy of the photon gas $S= (\partial F/\partial T)_{V}$ can now be written as 
\begin{eqnarray}
S &=&Nk_B\Big[4+\ln\Big\{\Big(\frac{4 \pi Vk_B^3T^3}{N}\Big)Q(\lambda,T)\Big\}\nonumber\\&&
-\Big\{
	60 e^{-\frac{1}{\lambda k_BT}}\Big(\frac{1}{\lambda^2 {k_B}^2 T^2}-\frac{2}{\lambda k_BT}+20-94\lambda k_BT-228\lambda^2  {k_B}^2 T^2
	\Big)\nonumber\\
	&&+24\lambda  k_BT-480\lambda^2  {k_B}^2 T^2\Big\}Q(\lambda,T)^{-1}\Big]+\mathcal{O}(\lambda^3)~.\nonumber\\
\label{entropy}
\end{eqnarray}
\noindent The relation between internal energy, free energy and entropy of this deformed photon gas $U = F + TS$, can be used to obtain, 
\begin{eqnarray}
U &= & Nk_B T\Big[3- \Big\{60 e^{-\frac{1}{\lambda k_BT}}\Big(\frac{1}{\lambda^2 {k_B}^2 T^2}-\frac{2}{\lambda k_BT}+20-94\lambda k_BT-228\lambda^2  {k_B}^2 T^2
	\Big)\nonumber\\&&+24\lambda  k_BT-480\lambda^2  {k_B}^2 T^2\Big\}Q(\lambda,T)^{-1}\Big]+\mathcal{O}(\lambda^3)~.
\label{internal energy}
\end{eqnarray} 
\normalsize As the density $ \rho = {U}/{V} $ is obtained from   internal energy and volume, we will have
\small{\begin{eqnarray}
P &=&  \Big(\frac{Nk_BT}{3V}\Big)Q(\lambda,T)^{-1}\Big[
	60 e^{-\frac{1}{\lambda k_BT}}\Big(\frac{1}{\lambda^2 {k_B}^2 T^2}-\frac{2}{\lambda k_BT}+20-94\lambda k_BT-228\lambda^2  {k_B}^2 T^2
	\Big)\nonumber\\
	&&24\lambda  k_BT-480\lambda^2  {k_B}^2 T^2\Big]+\frac{\rho}{3}+\mathcal{O}(\lambda^3)~.
\label{pressure-energy density relation}
\end{eqnarray}}
The  specific heat of this photon gas $C_v= (\partial U/\partial T)_{V}$ can be written as 
\begin{eqnarray}
C_v&=& Nk_B(3+C'Q(\lambda,T)^{-2})+\mathcal{O}(\lambda^3)\,,
\end{eqnarray}
where 
\begin{eqnarray}
C'&=&(-96\lambda k_B T + 3456 \lambda^2 {k_B}^2 T^2)-60 e^{-\frac{1}{\lambda k_BT}}\Big(\frac{2}{\lambda^3 {k_B}^3 T^3}-\frac{30}{\lambda^2 {k_B}^2 T^2}+\frac{400}{\lambda k_BT}-2452+9812\lambda k_BT\nonumber\\&&-34956\lambda^2  {k_B}^2 T^2\Big) +3600e^{-\frac{2}{\lambda k_BT}}\Big(-\frac{1}{\lambda^2 {k_B}^2 T^2}+80-1180\lambda k_BT+1084\lambda^2  {k_B}^2 T^2
\Big)+\mathcal{O}(\lambda^3)~.\nonumber\\
\end{eqnarray}
Figures~\ref{Fig1}-\ref{Fig3} represent the entropy, internal energy and specific heat versus the temperature plotting 
different dispersion models. In these plots, we have chosen the following values: $k_B=1, \lambda=10^{-4}, N=100000, V=0.01$. It may be noted that here we use $\lambda = 10^{-4}$ to study effects of such deformations, 
 and in the next section, we will discuss the  physically values for such a deformation. 
It may be noted that at low temperatures, the entropy of the photon gas is the same for special relativity (SR), the model used 
by Magueijo-Smolin (MS) and the modified dispersion relation (MDR) used in this paper. However, the entropy 
of MS is less than SR, and bigger than MDR at higher temperatures. The internal energy for SR is bigger than 
MDR and MS at high temperatures. However, the internal energy of MS is initially higher than MDR and then it becomes 
equal to MDR at about $T = 4000$. After that, the internal energy becomes smaller  than MDR. The specific heat in SR is constant and 
it is bigger than the specific heat for both MDR and MS models. The specific heat in MS is also constant for low temperature, and then it starts to 
reduce as the temperature increases. However, the specific heat for MS remains higher than MDR till  about $T= 2000$, then 
it becomes equal to   MDR at $T = 2000$. Finally, the specific heat becomes less than MDR. Hence, we observe that the thermodynamic properties 
depend critically on the modification of the usual energy-momentum dispersion relation.

\section{Dumb Hole}

In the previous section, we assumed the value $\lambda = 10^{-4}$. However, actually   
such a deformation  from quantum gravity is expected to occur at Planck scale, and for a deformation at such a scale, 
the deviation from the original   results at $T\sim 10^3$ would be very small. 
It is possible to assume such a deformation to the energy-momentum dispersion relation 
could occur at an intermediate scale between the Planck and
electroweak scales \cite{sa, sa12, de17}. In fact, the  Lorentz violation  breaking has been studied with 
 optical ring cavity, and it has been demonstrated that such a breaking can occur only at a scale $10^{-14}$ 
 \cite{1024}. 
 Even though this scale is still much larger than Planck scale, it is still too small to directly observe the 
 corrections to the thermodynamics from modified energy dispersion relation. So, we need a very sensitive   system to
 measure such effects. We will propose that dumb holes will form such a system.   A  dumb hole (also called as a sonic black hole), 
is an geometric object which traps  phonons (sound perturbations), just as a black hole traps photons   \cite{d1, d2}.
The geometry of a dumb hole depends on the   velocity of sound, so a deformation of the   velocity of sound will deform the 
geometry of a dumb hole. However, the trapping of sound by the horizon of the dumb hole is a very sensitive process, and even for a 
very small change in the velocity of sound, the  dumb hole would not form. Hence, if the velocity of sound changes even by a very small 
amount, the original horizon   will not be able to trap sound, and we will not get the dumb hole geometry. 
Therefore, we propose that dumb holes can be used to analyze the small differences in the velocity of sound produced by 
a deformation of  energy-momentum dispersion relation. 

\begin{figure}[H]
	\centering
	\includegraphics[scale=0.7]{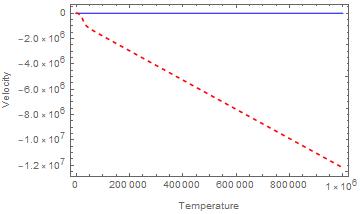}
	\caption{\textbf{Velocity of sound as a function of the Temperature $T$ for two cases}:
	Special Relativity (SR) and Modified Dispersion Relation (MDR). Here lower curve(red line) represents 
	MDR, upper curve(blue line) represents SR.}
	\label{Fig4}
\end{figure}

Now to demonstrate that, we will first have to analyze the effects of such a deformation of the  energy-momentum dispersion relation on
the velocity of sound.  So, first   we will calculate such effects   on 
the velocity of sound. It is known that the velocity of sound depends on the thermodynamics 
of photon gas  \cite{sound}, so that the velocity of sound should be deformed because of the deformation of 
thermodynamics. The velocity of  sound can be written as 
\begin{eqnarray}
v&=&\Big(\frac{E}{\rho}\Big)^{1/2}\,,\label{velocity}
\end{eqnarray}
where $\rho = U/V$, and the Young modulus $E$ is given by
\begin{eqnarray}
E&=&-V\frac{dP}{dV}\,. \label{Young}
\end{eqnarray} 
Here $V$ and $P$ are the volume and pressure of the photon gas. 
In our case, we have a relationship between $P$ and $V$ for this deformed photon gas given by
\begin{eqnarray}
P = \frac{\rho}{3}-\left(\frac{Nk_BT}{3V}\right)\alpha(\lambda,\beta)\,,\label{Prho}
\end{eqnarray}
where $\alpha$ is a long function of $\lambda$ and $\beta$ defined as
\begin{eqnarray}
\alpha(\lambda,\beta)&=&-\Big\{60 e^{-\frac{1}{\lambda k_BT}}\Big(\frac{1}{\lambda^2 {k_B}^2 T^2}-\frac{2}{\lambda k_BT}+20-94\lambda k_BT-228\lambda^2  {k_B}^2 T^2
	\Big)\nonumber\\
	&&+24\lambda  k_BT-480\lambda^2  {k_B}^2 T^2\Big\}Q(\lambda,T)^{-1}\,.
\end{eqnarray}
Thus, by using Eqs.~(\ref{Young}) and (\ref{velocity}), the Young modulus and the velocity of sound can be written as 
\begin{eqnarray}
E&=&\frac{\rho}{3}-\Big(\frac{Nk_{B}T}{3V}\Big)\alpha(\lambda,\beta)\,\label{Young2}\,,\\
v&=&\sqrt{\frac{1}{3}\Big[1-\frac{Nk_B T}{\rho V}\alpha(\lambda,\beta)\Big]}= \sqrt{\frac{1}{3}\Big[1-\frac{P}{\rho }\alpha(\lambda,\beta)\Big]}\,,
\label{velocity-s}
\end{eqnarray}
where in the second equation we have used $PV=Nk_{B}T$. As  $v$ depends on $\lambda$ and $\beta$, we will write $v$ as 
 $v(\lambda,\beta)$. This will be important as  we will need this to show that the dumb hole horizon also depends on $\lambda$ and $\beta$. \\
By using Eq.~(\ref{Prho}) (and again  using $PV=Nk_{B}T$), we have that the pressure is
\begin{eqnarray}
P=\frac{\rho}{3+\alpha(\lambda,\beta)}\,.\label{press3z}
\end{eqnarray}
Using the above relation in Eq.~(\ref{velocity-s}), we finally obtain that the velocity of sound is given by
\begin{eqnarray}
v(\lambda,\beta)=\frac{1}{\sqrt{3+\alpha(\lambda,\beta)}}\,.
\label{velocity3}
\end{eqnarray} 
This is the expression for the modified velocity of sound in photon gas. It may be noted that this modification of the velocity 
of sound is due to the deformation of the energy-momentum dispersion relation and it will have interesting astrophysical 
application, as it will effect the accretion  of   CMB photons around these primordial black holes \cite{accre, accre1}. It is reassuring to note that
the above expression reduces to the usual result in the $\lambda\rightarrow 0$
limit \cite{sound}
\begin{eqnarray}
v=\frac{1}{\sqrt{3}}\,.
\label{usualvelocity3}
\end{eqnarray}
Remark that the velocity of sound depends critically on the temperature due to a deformation of the usual energy-momentum dispersion 
relation. Such a modification to the velocity  of sound becomes significant at higher temperatures. Further, the velocity of sound 
is constant when the usual energy-momentum dispersion relation is used. Thus, we can use this modified expression for the velocity 
of sound to analyse the physical effects of the deformed energy-momentum dispersion relation. Fig.~\ref{Fig4} shows the velocity of sound versus Temperature for Special Relativity (which assumes the standard energy-momentum dispersion) and MDR.

Now after we have analysed the effect of the modified dispersion relation on the velocity of sound, we can analyse the effect of 
such a deformation on dumb holes \cite{d1, d2}. 
The  deformed energy-momentum tensor and the current for a perfect fluid, can be written as 
\begin{eqnarray}
	\label{Tmunu} 
	T^{\mu\nu} &=& Pg^{\mu\nu} + (\rho+P)u^\mu u^\nu\nonumber \\ &=&
	\frac{\rho}{3+\alpha(\lambda,\beta)}\Big[g^{\mu\nu}+(4+\alpha(\lambda,\beta)) u^\mu u^\nu\Big] \,, \\
	\label{current}
	j_i^\mu &=& q_i u^\mu\,, 
\end{eqnarray}
where $q_{i}$ are the conserved charges and $u^{\mu}$ is the four velocity which satisfies $u^{\mu}u_{\mu}=-1$.
It may be noted that this four velocity satisfies a similar relation to the original four velocity \cite{Das:2010mk}. 
These two quantities are conserved as $\nabla_{\mu}T^{\mu\nu}=0$ and $\nabla_{\mu}j^{\mu}_{i}=0$. Now, we can define  
\begin{eqnarray}
\bar{\mathcal{T}}^{4}= \frac{\rho}{3+\alpha(\lambda,\beta)}=P\,.
\end{eqnarray}
Clearly, if we set $\alpha=0$, the standard equation of state is recovered. Moreover, if one assumes that $\alpha(\lambda,\beta)\ll 1$, 
and one expands the above expression up to first order in $\alpha$, one gets
\begin{eqnarray}
P&=&\frac{\rho }{3}-\frac{\alpha  \rho }{9}+\mathcal{O}(\alpha^2)\nonumber \\ &=& 
P_{0}-\frac{\alpha  \rho }{9}+\mathcal{O}(\alpha^2)\nonumber \\ 
&=& \mathcal{T}^{4}-\frac{\alpha  \rho }{9}+\mathcal{O}(\alpha^2)\,,
\end{eqnarray}
where $P_{0}=\rho/3$ and $\mathcal{T}^{4}=P_{0}$ are    the quantities computed in the standard theory  which assumes the usual energy-momentum dispersion.
Therefore, up to first order in $\alpha$, we have that the pressure differs by  $-\alpha  \rho/9$.

We can project the  energy-momentum conservation law onto the four velocity as $u_{\nu}\nabla_{\mu}T^{\mu\nu}=0$,  yielding
\begin{eqnarray}
	\nabla_{\mu}\Big(\bar{\mathcal{T}}^{3}u^{\mu}\Big)=0\,.
\end{eqnarray}
Now, by projecting the energy-momentum tensor onto the projector $h^{\lambda}_{\nu}=\delta^{\lambda}_{\nu}+u^{\lambda}u_{\nu}$, we obtain 
\begin{eqnarray}
	\nabla_{\mu}\Big(\bar{\mathcal{T}}u_{\nu}\Big)&=&\nabla_{\nu}\Big(\bar{\mathcal{T}}u_{\mu}\Big)\,,
\end{eqnarray}
where we have assumed that the fluid satisfies $h^{\lambda}_{\mu}h^{\sigma}_{\nu}(\partial_{\lambda}u_{\sigma}-\partial_{\sigma}u_{\lambda})\equiv0$. 
This kind of fluids are called irrational since they do not have a vorticity term. The above equation allows us to define a potential such as
\begin{eqnarray}
\bar{\mathcal{T}}u_{\mu}=\partial_{\mu}\phi\,,\label{35}
\end{eqnarray}
and hence by using $u_{\mu}u^{\mu}=-1$,  we also get
\begin{eqnarray}
	\bar{\mathcal{T}}^2=-(\partial_{\mu}\phi)(\partial^{\mu}\phi)\,.\label{36}
\end{eqnarray}
We have all the ingredients to study isentropic sound waves for this kind of fluid. 
To do this, we need to take small variations of the velocity potential $\phi\rightarrow \phi+\delta\phi$.
By doing that, all our other quantities also are deformed as follows
\begin{eqnarray}
	u^{\mu}&\rightarrow& u^{\mu}+\delta u^{\mu}\,,\\
	q_{i}&\rightarrow& q_{i}+\delta q_{i}\,,\\
\bar{\mathcal{T}}&\rightarrow& \bar{\mathcal{T}}+\delta \bar{\mathcal{T}}\,.
\end{eqnarray}
If we replace these quantities in (\ref{35}), we find the following expression
\begin{eqnarray}
u^{\mu}\delta \bar{\mathcal{T}}+\bar{\mathcal{T}}\delta u^{\mu}&=&\partial^{\mu}\delta \phi\,.
\end{eqnarray}
Now, if we use that $u_{\mu}\delta u^{\mu}=0$,  we obtain 
$\bar{\mathcal{T}}\delta u^{\mu}=h^{\mu\nu}\partial_{\nu}(\delta \phi)$,
and $\delta \bar{\mathcal{T}}=-u^{\mu}\partial_{\mu}\delta \phi$. By using these expressions in the above equation, we obtain 
\begin{eqnarray}\label{phi}
	\partial_{\mu}\Big[\sqrt{-g}\bar{\mathcal{T}}^2(g^{\mu\nu}-2u^{\mu}u^{\nu})\partial_{\nu}\Big](\delta \phi)=0\,.
\end{eqnarray} 
Now we can write  the acoustic metric as 
\begin{eqnarray}
G_{\mu\nu}=\sqrt{3}\bar{\mathcal{T}}^{2}(g_{\mu\nu}+\frac{2}{3}u_{\mu}u_{\nu})\,,
\end{eqnarray}
and then we can rewrite Eq.~(\ref{phi}) as a Klein-Gordon massless scalar field equation, yielding
\begin{eqnarray}	
\partial_{\mu}\Big[\sqrt{-G}G^{\mu\nu}\partial_{\nu}\Big](\delta \phi)=0\,.
\end{eqnarray}
The metric $G^{\mu\nu}$ describes the acoustic metric of the dumb hole, and such a metric has been constructed for a dumb hole with the usual 
energy-momentum dispersion relation \cite{Das:2010mk}. It may be noted that it is possible to construct a similar metric using the 
deformed energy-momentum dispersion relation, and changing the 
 the time-coordinate as \begin{eqnarray}
                         d\tau=dt+\frac{\frac{3}{2}\gamma(\lambda,\beta)^2v_{i}}{1-\frac{2}{3}\gamma(\lambda,\beta)^2}dx^{i}\,,    
                         \end{eqnarray}
where $\gamma(\lambda,\beta)$ is the Lorentz like 
factor for this metric, in which the velocity of light is replaced by the velocity of sound 
\cite{Das:2010mk}. It may be noted that for the deformed case, the velocity of sound is a function of $\lambda$
and $\beta$, and $\gamma$ is the Lorentz like factor with velocity of sound as the limiting velocity, 
so the deformed $\gamma$ will also be a function of $\lambda$
and $\beta$. 
Thus, following the calculations done for the original dumb holes  \cite{Das:2010mk}, 
the corresponding metric for the dumb holes in the deformed theory can be written as 
\begin{eqnarray}
	\label{dst2}
	d\tilde{s}^2 = \sqrt{3}\bar{\mathcal{T}}^{2}\left\{ -\Big(1-\frac{2}{3}\gamma(\lambda,\beta)^2\Big)d\tau^2 +
	\left( g_{ij} +
	\frac{\frac{2}{3}\gamma(\lambda,\beta)^2}{1-\frac{2}{3}\gamma(\lambda,\beta)^2}v_iv_j\right)
	dx^idx^j \right\}\,. \end{eqnarray}
 Now  this metric has a horizon at $\gamma(\lambda,\beta) = \sqrt{3/2}$, and 
 this horizon also is a function of $ \lambda $ and $ \beta$. The horizon of a dumb hole by definition traps the sound. 
 However, as the velocity of sound is a function of $ \lambda $ and $ \beta$, the original horizon is not the horizon 
 in the deformed theory, as it cannot trap the sound in the deformed theory. Now for a given value of $\lambda$, 
 the horizon is a function of temperature, as the sound is also a function of temperature. However, for a given temperature 
 and a given value of $\lambda$, this theory has a well defined  fixed horizon. 
So, the geometry of the dumb holes  get deformed due to a deformation of the energy-momentum dispersion relation.
Such a deformation of the geometry 
 of dumb holes can have observational consequences, and it would be interesting to analyse such consequences. 
 In fact, one such observational consequence occurs due to the  analogous Hawking radiation from a 
 dumb hole. 
 This dumb hole can emit Hawking radiation at
 a temperature   \cite{hawk, hawk1}
 \begin{eqnarray}
  T = \frac{\kappa}{2\pi }\,, 
 \end{eqnarray}
 where $\kappa$ is the analogous surface gravity of this dumb hole. 
 It would be possible to measure this analogous Hawking radiation and observe at what values the dumb hole form, as 
  the Hawking radiation will only occur when a dumb hole forms. 
 However,  formation of the  dumb hole gets modified by the deformation  of the energy-momentum dispersion relation, 
 so we can use the Hawking radiation to measure this deformation.  In fact, as the  formation of 
 a dumb hole is very sensitive to even   small   changes in the velocity of sound, 
 this system can detect such small changes in the velocity of sound, which might not be observed by various other effects.  
 So, a dumb hole can be used to measure deformation of the usual energy-momentum dispersion relation.

\section{Conclusion}

In this paper, we have analysed the deformation of the usual energy-momentum dispersion relation. 
We deformed the usual energy-momentum dispersion relation at an intermediate scale between the Planck and electroweak scales. 
We have investigated the  effect of such a deformation on the   physics of a  photon gas. It was
 observed that the deformation of the usual energy-momentum dispersion relation will deform the partition function for 
the photon gas. It was also demonstrated that the  deformation of the partition function deformed the thermodynamics of the photon gas. Therefore, we
performed a derivation of the corrections 
to various thermodynamic quantities from the deformation of the usual energy-momentum dispersion relation. 
The analysis we performed in this paper was more general than the   previous works, where high temperature limit was taken \cite{hight}. 
Note that since we were interested on studying dumb holes and they are usually constructed using Bose-Einstein condensates 
  \cite{cobe, cobe1}, and such condensates can become unstable at high temperature, we had to analyse the modification to the thermodynamics without 
  taking such a high temperature limit. 
 Furthermore, in this work a different form of modified dispersion relation (MDR) was used than the 
 MS model. It may be noted that these corrections only become significant in the UV limit of the theory, and the usual 
thermodynamics is recovered in the IR limit of the theory. This is because the corrections depend on the factor 
$\lambda$, and in the IR  limit $\lambda \rightarrow 0$, as in the IR limit 
the deformed energy-momentum dispersion relation reduces to the usual energy-momentum 
dispersion relation. Thus, in the IR limit, the correction terms for the modified  thermodynamics vanish, and we recover the 
usual thermodynamics. However, the correction terms cannot be neglected in the UV limit  of this theory. 
As the velocity of sound depends on the thermodynamics of 
the photon gas \cite{sound}, a deformation of this thermodynamics  also modified the expression for the velocity of sound. 
In this paper, we have explicitly  observed the effect of deforming the usual energy-momentum dispersion relation on
the velocity of sound.  
Furthermore, as the horizon of a dumb hole depends on the local velocity of sound, a change in the local velocity of sound would change 
the horizon of a dumb hole.   We have thus analysed the effect of the deformation of the usual energy-momentum dispersion relation 
on the geometry of a dumb hole. This deformation of the horizon of a dumb hole can also deform the Hawking radiation 
 and this deformation in the Hawking radiation can be detected. This can be used to test the breaking of the Lorentz 
 symmetry at a scale between the electroweak 
and Planck scales. Thus, in this paper, it has been argued that the 
 geometry of a dumb hole is non-trivially deformed by the deformation 
of the usual energy-momentum dispersion relation, and dumb holes can be in turn  used to
test such deformations of the usual energy-momentum dispersion relation. 

The velocity of sound fixes the value of the sonic points, and the  sonic 
 points are important for many important process like 
such  accretion around  a  black hole    \cite{1u,W1,W2,2u}.  
The accretion around a  black hole was first studied    by Bondi in the Newtonian framework \cite{4u}, and so it is called 
 Bondi   accretion. 
 The relativistic generalization of  the Bondi type accretion  has   been used for analysing  accretion  
around  a   black hole \cite{5u,6u}.  It has been demonstrated that 
 radiative processes can have interesting effects for accretion around a black hole 
 \cite{7z,9z,z9} or other spherically symmetric space-times \cite{Bahamonde:2015uwa}. The accretion also gets effected from   rotation  of a black hole \cite{8z}.
 It has been observed that   the cosmological constant can also effect the  accretion  around a black hole \cite{1t, t}.
It has also been demonstrated that there is 
  a correspondence between sonic points of ideal photon gas and photon spheres, and this correspondence has
  been used to study the accretion   of ideal photon gas
\cite{phot}. In fact,  it is possible for that the primordial black holes would be formed at early stages of the evolution 
of this universe, and the   photon gas of CMB photon would couple to such 
 primordial black holes. Therefore, the accretion 
 of photons would occur around such   primordial black holes. In fact,   such accretion of CMB photons around 
 primordial black holes has already been discussed \cite{accre, accre1}.
 In this analysis, it was observed that the 
 accretion  of   CMB photons around these primordial black holes  
depends on the velocity of sound. This is because the velocity of sound 
fixes the value of the sonic points, and the 
 accretion of the CMB photons depend on the transonic solution 
 across the sonic point. As it was demonstrated 
 in this paper, that the velocity of sound changes due to a deformation of the usual energy-momentum dispersion relation,  
 so, the  results of this paper can have direct application for analysing 
 the accretion  of   CMB photons around these primordial black holes.


\section*{Acknowledgments}
S.G. acknowledges the support by DST SERB,  India under Start Up Research Grant (Young Scientist), File No.YSS/2014/000180. S.B. is supported by the Comisi\'on Nacional de Investigaci\'on
Cient\'ifica y Tecnol\'ogica (Becas Chile Grant No. 72150066).  The authors would also like to thank the referee for very useful comments.



\begin{thebibliography}{99} 
 
\bibitem{Iengo} R. Iengo, J. G. Russo and M. Serone, JHEP   {11}, 020
(2009).

\bibitem{Arkani} A. Adams, N. Arkani-Hamed, S. Dubovsky, A. Nicolis and R.
Rattazzi, JHEP   {10}, 014 (2006).

\bibitem{Gripaios} B. M. Gripaios, JHEP   {10}, 069 (2004).

\bibitem{Alfaro} J. Alfaro, P. Gonzalez and R. Avila, Phys. Rev. D   {91%
}, 105007 (2015).

\bibitem{Belich} H. Belich and K. Bakke, Phys. Rev. D   {90}, 025026
(2014).
 

\bibitem{Hooft} G. 't Hooft, Class. Quantum Gravit.  {13}, 1023 (1996).
 

\bibitem{Ellis} G. Amelino-Camelia, J. R. Ellis, N. Mavromatos, D. V.
Nanopoulos and S. Sarkar, Nature  {393}, 763 (1998).

\bibitem{Gambini} R. Gambini and J. Pullin, Phys. Rev. D  {59}, 124021
(1999).

\bibitem{FaizalMPLA} M. Faizal, Mod. Phys. Lett. A  {27}, 1250075
(2012).

\bibitem{Carroll} S.~M. Carroll, J.~A. Harvey, V.~A. Kostelecky, C.~D. Lane
and T.~Okamoto, Phys. Rev. Lett.  {87}, 141601 (2001).

\bibitem{FaizalJPA} M. Faizal, J. Phys. A  {44}, 402001 (2011).

\bibitem{h1} P.~Horava, Phys. Rev. D  {79}, 084008 (2009).

\bibitem{h2} P.~Horava, Phys. Rev. Lett.  {102}, 161301 (2009).

\bibitem{Bahamonde:2015zma}
S.~Bahamonde, C.~G.~Böhmer and M.~Wright,
Phys.\ Rev.\ D {\bf 92} (2015) no.10,  104042

\bibitem{Bahamonde:2017bps}
S.~Bahamonde, S.~Capozziello, M.~Faizal and R.~C.~Nunes,
Eur.\ Phys.\ J.\ C {\bf 77} (2017) no.9,  628

\bibitem{Bahamonde:2016kba}
S.~Bahamonde and C.~G.~Böhmer,
Eur.\ Phys.\ J.\ C {\bf 76} (2016) no.10,  578

\bibitem{Bahamonde:2015hza}
S.~Bahamonde and M.~Wright,
Phys.\ Rev.\ D {\bf 92} (2015) no.8,  084034
Erratum: [Phys.\ Rev.\ D {\bf 93} (2016) no.10,  109901]

\bibitem{13} J . Magueijo, and L. Smolin   Phys. Rev. D   71, 026010 (2005). 



  \bibitem{gr} J. Magueijo and L. Smolin, Class. Quantum Gravit.  {%
21}, 1725 (2004).


\bibitem{58} V. A. Kostelecky and S. Samuel, Phys. Rev. D   {39}, 683
(1989).

\bibitem{59} V. A. Kostelecky and S. Samuel, Phys. Rev. D   {40}, 1886
(1989).

\bibitem{60} S. Mukohyama, JHEP   {05}, 048 (2007).

\bibitem{61} P. West, Phys. Lett. B   {548}, 92 (2002).


\bibitem{sa} S.  Das and  E. C. Vagenas,  Phys. Rev. Lett. 101, 221301 (2008)
\bibitem{sa12}  A. F.  Ali, S.  Das and  E. C. Vagenas, Phys. Rev. D 84, 044013 (2011)  

\bibitem{md01}G. Amelino-Camelia, M. Arzano, Y. Ling and G. Mandanici, Class. Quant. Grav. 23,
2585 (2006).
\bibitem{md02}L.  Xiang, Phys. Lett. B 638, 519 (2006).




\bibitem{Greisen} K. Greisen, Phys. Rev. Lett.  {16}, 748 (1966).

\bibitem{Zatsepin} G. T. Zatsepin and V. A. Kuzmin, JETP Lett.  {4},
78 (1966).

\bibitem{Abraham} J. Abraham et al. (Pierre Auger Collaboration), Phys.
Lett. B  {685}, 239 (2010).

 


  \bibitem{AmelinoCamelia:1997gz}
  G.~Amelino-Camelia, J.~R.~Ellis, N.~E.~Mavromatos, D.~V.~Nanopoulos and S.~Sarkar,
   Nature {  393}, 763 (1998). 
   
    \bibitem{sound} M. Pardy,  Results  Phys. 3,  70 (2013).
   \bibitem{hight}  S. H. S. Alexander and J. Magueijo,    Proc. of the XIIIrd Rencontres de Blois, 
Frontiers of the Universe, 281  (2004).
    \bibitem{ther2} S.  Das,  S.  Ghosh  and  D.  Roychowdhury,  Phys. Rev. D80, 125036 (2009).
 \bibitem{ther}  S. Das, S. Pramanik and S. Ghosh,  SIGMA 10, 104 (2014). 
 
 \bibitem{ther1}  S. Das and  D. Roychowdhury, Phys. Rev. D 81, 085039 (2010).

 
  \bibitem{ther0}  X. Zhang, L. Shao and B-Q. Ma,  Astropart. Phys. 34, 840 (2011).  
  
  \bibitem{d1}M. Visser, Class. Quant. Grav. 15, 1767 (1998). 
  \bibitem{d2} J. Steinhauer,  Nature Phys.     10, 864 (2014).
 
  
     \bibitem{cobe} L. J. Garay, J. R. Anglin, J. I. Cirac and  P. Zoller, Phys. Rev. Lett. 85, 4643 (2000) 
  \bibitem{cobe1} L. J. Garay, J. R. Anglin, J. I. Cirac and P. Zoller,  Phys. Rev. A 63, 023611 (2001)
  \bibitem{conden} O. Lahav, A. Itah, A. Blumkin, C. Gordon, S.  Rinott, A.  Zayats and J.  Steinhauer, 
  Phys. Rev. Lett. 10, 240401 (2010)
  
  
   \bibitem{hawk} S. Giovanazzi, Phys. Rev. Lett. 94, 061302 (2005)
  
  \bibitem{Das:2010mk}
  S.~R.~Das, A.~Ghosh, J.~H.~Oh and A.~D.~Shapere,
  JHEP {  1104} 030 (2011). 
  
  \bibitem{Magueijo-smolin}J. Magueijo and L. Smolin, Phys. Rev. Lett. 88, 190403 (2002).
  \bibitem{Magueijo-smolin1}J. Magueijo and L. Smolin, Phys. Rev. D 67, 044017 (2003). 
  
  
   \bibitem{Alae_Brand_Mag} S. Alexander, R. Brandenberger and J. Magueijo, Phys. Rev. D 67, 081301 (2003).
  \bibitem{amelino-camelia-Mazid} G. Amelino-Camelia and S. Majid, Int. J. Mod. Phys. A 15, 4301  (2000).
  
 
  
 
  \bibitem{de17}G. Amelino-Camelia, C. Laemmerzahl, F. Mercati and G. M. Tino,  Phys. Rev. Lett. 103, 171302 (2009)
   \bibitem{1024}  Y. Michimura, N. Matsumoto, N. Ohmae, W. Kokuyama, Y. Aso, M. Ando and K. Tsubono,
  Phys. Rev. Lett. 110, 200401 (2013)
  
  
  \bibitem{hawk1} D. Boiron, A. Fabbri, P. E. Larre, N. Pavloff, C. I. Westbrook and  P. Zin, Phys. Rev. Lett. 115, 025301 (2015)
 
  
  
  
  
  
   \bibitem{1u} S.K. Chakrabarti, Phys. Rept. 266, 229 (1996).
  \bibitem{W1}T. Matsuda, M. Inoue and K. Sawada, Mon. Not. R. Astron. Soc. 226, 785  (1987).
  \bibitem{W2}R. Taam and B. Fryxall, Astrophys. J. 331, L117 (1988).
  \bibitem{2u} E. Shima et al. Mon. Not. R. Astron. Soc. 217, 367 (1985).
  
  \bibitem{4u} H. Bondi, Mon. Not. Roy. Astron. Soc. 112, 195 (1952).
  
  \bibitem{5u} F.C. Michel, Astrophys. Space Sci. 15, 153 (1972).
  \bibitem{6u}K. S. Thorne, R.A. Flammang and A.N. Zytkow, Mon. Not. R. Astron. Soc. 194, 475 (1981).
  
  \bibitem{7z}S. L. Shapiro, Astrophys. J. 185, 69 (1973).
  \bibitem{9z}S. L. Shapiro, Astrophys. J. 189, 343 (1974).
  \bibitem{z9} W. Brinkmann, Astron. Astrophys.
  85, 146 (1980).
  
  \bibitem{Bahamonde:2015uwa}
  S.~Bahamonde and M.~Jamil,
  Eur.\ Phys.\ J.\ C {\bf 75} (2015) 508
  
  
  \bibitem{8z}G.R. Blumenthal and W.G. Mathews, Astrophys. J. 203, 714 (1976).
  
  
  \bibitem{1t} J. Karkowski, E. Malec,  Phys. Rev. D
  87, 044007
  (2013).
  \bibitem{t} P.  Mach and   E.  Malec,
  Phys. Rev. D
  88, 084055 (2013).
  
  
  
 \bibitem{phot} Y. Koga and  T. Harada, Phys. Rev. D 94, 044053 (2016).



 

\bibitem{accre}  M. Ricotti,  	Astrophys. J. 662, 53 (2007).
 \bibitem{accre1}J. A. Fillmore, and P. Goldreich,   Astrophys. J.  281, 1 (1984). 
 

 


 







 
\end{thebibliography}
\end{document}